\documentclass[12pt]{article}
\usepackage{amsmath,amssymb,amsthm,amsxtra,overpic,bbm,bm,epsfig,subfigure}
\usepackage{color}
\usepackage{comment}
\usepackage[all,cmtip]{xy}

\usepackage{slashed,stmaryrd}

\renewcommand{\thefootnote}{\fnsymbol{footnote}}

\begin{document}

\vspace{0.2cm}

\begin{center}
{\large\bf Symmetric formulation of neutrino oscillations in matter and \\ its intrinsic connection to renormalization-group equations}
\end{center}

\vspace{0.2cm}

\begin{center}
{\bf Shun Zhou} \footnote{E-mail: zhoush@ihep.ac.cn}
\\
{Institute of High Energy Physics, Chinese Academy of
Sciences, Beijing 100049, China \\
Center for High Energy Physics, Peking University, Beijing 100871, China}
\end{center}

\vspace{1.5cm}

\begin{abstract}
In this article, we point out that the effective Hamiltonian for neutrino oscillations in matter is invariant under the transformation of the mixing angle $\theta^{}_{12} \to \theta^{}_{12} - \pi/2$ and the exchange of first two neutrino masses $m^{}_1 \leftrightarrow m^{}_2$, if the standard parametrization of lepton flavor mixing matrix is adopted. To maintain this symmetry in perturbative calculations, we present a symmetric formulation of the effective Hamiltonian by introducing an $\eta$-gauge neutrino mass-squared difference $\Delta^{}_* \equiv \eta \Delta^{}_{31} + (1-\eta)\Delta^{}_{32}$ for $0 \leq \eta \leq 1$, where $\Delta^{}_{ji} \equiv m^2_j - m^2_i$ for $ji = 21, 31, 32$, and show that only $\eta = 1/2$, $\eta = \cos^2\theta^{}_{12}$ or $\eta = \sin^2 \theta^{}_{12}$ is allowed. Furthermore, we prove that $\eta = \cos^2 \theta^{}_{12}$ is the best choice to derive more accurate and compact neutrino oscillation probabilities, by implementing the approach of renromalization-group equations. The validity of this approach becomes transparent when an analogy is made between the parameter $\eta$ herein and the renormalization scale $\mu$ in relativistic quantum field theories.
\end{abstract}

\begin{flushleft}
\hspace{0.88cm} PACS number(s): 14.60.Pq, 25.30.Pt
\end{flushleft}

\def\thefootnote{\arabic{footnote}}
\setcounter{footnote}{0}

\newpage

{\bf Introduction} --- Neutrino oscillation experiments in the last few decades have provided us with compelling evidence for tiny neutrino masses and significant lepton flavor mixing. This great achievement in elementary particle physics has been recognized by the Nobel Prize in Physics in 2015~\cite{Kajita:2016cak,McDonald:2016ixn}. In the framework of three neutrino flavors, lepton flavor mixing can be described by a $3 \times 3$ unitary matrix $U$, i.e., the Pontecorvo-Maki-Nakagawa-Sakata (PMNS) matrix~\cite{Pontecorvo:1957cp,Maki:1962mu}, which is usually parametrized in terms of three mixing angles $\{\theta^{}_{12}, \theta^{}_{13}, \theta^{}_{23}\}$ and one CP-violating phase $\delta$. Adopting the standard parametrization advocated by the Particle Data Group~\cite{Olive:2016xmw}, we have
\begin{eqnarray}
U = R(\theta^{}_{23})\cdot R(\theta^{}_{13}, \delta) \cdot R(\theta^{}_{12}) \equiv \left(\begin{matrix} 1 & 0 & 0 \cr
0 & c^{}_{23} & s^{}_{23} \cr 0 & -s^{}_{23} & c^{}_{23}\end{matrix}\right)\left(\begin{matrix} c^{}_{13} & 0 & s^{}_{13}e^{-{\rm i}\delta} \cr
0 & 1 & 0 \cr -s^{}_{13}e^{{\rm i}\delta} & 0 & c^{}_{13}\end{matrix}\right)\left(\begin{matrix} c^{}_{12} & s^{}_{12} & 0 \cr -s^{}_{12} & c^{}_{12} & 0 \cr 0 & 0 & 1\end{matrix}\right) \; , ~~
\label{eq:pmns}
\end{eqnarray}
where $c^{}_{ij} \equiv \cos\theta^{}_{ij}$ and $s^{}_{ij} \equiv \sin \theta^{}_{ij}$ have been defined for $ij = 12, 13, 23$, $R(\theta^{}_{ij})$ denotes a rotation matrix in the $i$-$j$ plane with a rotation angle $\theta^{}_{ij}$, and $R(\theta^{}_{13}, \delta) = U^{}_\delta R(\theta^{}_{13}) U^\dagger_\delta$ with $U^{}_\delta \equiv {\rm diag}\{1, 1, e^{{\rm i}\delta}\}$. At present, three mixing angles $\theta^{}_{12} \approx 34^\circ$, $\theta^{}_{23} \approx 45^\circ$ and $\theta^{}_{13} \approx 9^\circ$, together with two neutrino mass-squared differences $\Delta^{}_{21} \equiv m^2_{2} - m^2_1 \approx 7.5\times 10^{-5}~{\rm eV}^2$ and $|\Delta^{}_{31}| \equiv |m^2_3 - m^2_1| \approx 2.5\times 10^{-3}~{\rm eV}^2$, have been well determined from neutrino oscillation experiments~\cite{Olive:2016xmw}. The primary goals of future experiments are to pin down neutrino mass ordering, i.e., the sign of $\Delta^{}_{31}$, and to probe the leptonic CP-violating phase $\delta$.

To achieve these goals, the ongoing and forthcoming oscillation experiments are designed for medium- or long-baseline lengths, and neutrino beams are actually propagating through the Earth. In this case, the impact of a coherent forward scattering of neutrinos with background electrons can be taken into account by an effective matter potential $V = \sqrt{2}G^{}_{\rm F} N^{}_e$, where $G^{}_{\rm F} = 1.167\times 10^{-5}~{\rm GeV}^{-2}$ is the Fermi constant and $N^{}_e$ stands for the net electron number density. It is well known that the matter potential can dramatically modify neutrino flavor conversions~\cite{Wolfenstein:1977ue,Mikheev:1986gs}. For antineutrinos, the matter potential will change to a minus sign. Considering a neutrino beam of energy $E$ travelling in matter, we can write down the effective Hamiltonian for neutrino flavor oscillations~\cite{Kuo:1989qe,Xing:2003ez,Blennow:2013rca}
\begin{eqnarray}
\widetilde{H}^{}_{\rm eff} = \frac{1}{2E} \left[U \left(\begin{matrix} m^2_1 & 0 & 0 \cr 0 & m^2_2 & 0 \cr 0 & 0 & m^2_3 \end{matrix}\right) U^\dagger + \left(\begin{matrix} A & 0 & 0 \cr 0 & 0 & 0 \cr 0 & 0 & 0 \end{matrix}\right)\right] \equiv \frac{\widetilde{\Omega}^{}_\nu}{2E}\; ,
\label{eq:hami}
\end{eqnarray}
with $A \equiv 2EV$ and $\widetilde{\Omega}^{}_\nu$ being defined as the square of the effective neutrino mass matrix in matter. As usual, one can diagonalize the effective Hamiltonian by the corresponding PMNS matrix $\widetilde{U}$ in matter, namely,
\begin{eqnarray}
\widetilde{\Omega}^{}_\nu = \widetilde{U} \left(\begin{matrix} \widetilde{m}^2_1 & 0 & 0 \cr 0 & \widetilde{m}^2_2 & 0 \cr 0 & 0 & \widetilde{m}^2_3 \end{matrix}\right) \widetilde{U}^\dagger \; ,
\label{eq:omega}
\end{eqnarray}
where $\widetilde{m}^{}_i$ for $i = 1, 2, 3$ are neutrino masses in matter and $\widetilde{U}$ can be parametrized in terms of effective mixing parameters $\{\widetilde{\theta}^{}_{12}, \widetilde{\theta}^{}_{13}, \widetilde{\theta}^{}_{23}\}$ and $\widetilde{\delta}$ in the same way as $U$ in Eq.~(\ref{eq:pmns}).

With the help of three effective neutrino masses $\widetilde{m}^{}_i$ and the flavor mixing matrix $\widetilde{U}$, it is straightforward to calculate neutrino oscillation probabilities for a constant matter density~\cite{XZ}.
Moreover, based on the structure of $\widetilde{H}^{}_{\rm eff}$ and its relation to the Hamiltonian in vacuum, one can derive the Naumov relation $\widetilde{J} \widetilde{\Delta}^{}_{21} \widetilde{\Delta}^{}_{31} \widetilde{\Delta}^{}_{32} = J \Delta^{}_{21} \Delta^{}_{31} \Delta^{}_{32}$~\cite{Naumov:1991ju,Harrison:1999df,Xing:2000gg,Xing:2000ik}, where $\widetilde{J}$ and $J$ are respectively the Jarlskog invariants in matter and in vacuum~\cite{Jarlskog:1985ht}, and also obtain the Toshev relation $\sin 2\widetilde{\theta}^{}_{23} \sin \widetilde{\delta} = \sin 2\theta^{}_{23} \sin\delta$~\cite{Toshev:1991ku,Kimura:2002wd}. These identities are very useful in understanding the relationship between matter-corrected mixing parameters and the intrinsic ones.

\vspace{0.5cm}

{\bf Symmetric formulation} --- In practice, it is necessary to express the oscillation probabilities in terms of $\{\theta^{}_{12}, \theta^{}_{13}, \theta^{}_{23}, \delta\}$ and $\{\Delta^{}_{21}, \Delta^{}_{31}\}$, which are the fundamental parameters to be extracted from oscillation experiments. To this end, we can follow a direct diagonalization of $\widetilde{H}^{}_{\rm eff}$ or equivalently $\widetilde{\Omega}^{}_\nu$ and calculate the eigenvalues and eigenvectors. Then, the derived exact oscillation probabilities can be expanded in terms of some small parameters. Before doing so, we should first explore the basic properties of the effective Hamiltonian, by recasting $\widetilde{\Omega}^{}_\nu$ into the following form
\begin{eqnarray}
\widetilde{\Omega}^0_\nu = \left(\begin{matrix} c^{}_{12} & s^{}_{12} & 0 \cr -s^{}_{12} & c^{}_{12} & 0 \cr 0 & 0 & 1\end{matrix}\right) \left(\begin{matrix} m^2_1 & 0 & 0 \cr 0 & m^2_2 & 0 \cr 0 & 0 & m^2_3\end{matrix}\right) \left(\begin{matrix} c^{}_{12} & -s^{}_{12} & 0 \cr s^{}_{12} & c^{}_{12} & 0 \cr 0 & 0 & 1\end{matrix}\right) + A \left(\begin{matrix} c^2_{13} & 0 & c^{}_{13} s^{}_{13} \cr 0 & 0 & 0 \cr c^{}_{13} s^{}_{13} & 0 & s^2_{13}\end{matrix}\right) \; ,
\label{eq:omega0}
\end{eqnarray}
where the standard parametrization in Eq.~(\ref{eq:pmns}) is taken and a unitary transformation in the flavor space $\widetilde{\Omega}^0_\nu = [R^\dagger(\theta^{}_{13}) \cdot U^\dagger_\delta \cdot R^\dagger(\theta^{}_{23})] \cdot \widetilde{\Omega}^{}_\nu \cdot [R(\theta^{}_{23}) \cdot U^{}_\delta \cdot R(\theta^{}_{13})]$ is performed. The fact that $U^\dagger_\delta$ and $U^{}_\delta$ commute with $R(\theta^{}_{12})$ and the diagonal matrix ${\rm diag}\{m^2_1, m^2_2, m^2_3\}$ should be noted as well.

The transformation in the flavor space by a unitary matrix $\hat{U} \equiv R(\theta^{}_{23}) \cdot U^{}_\delta \cdot R(\theta^{}_{13})$ does not affect the eigenvalues of $\widetilde{\Omega}^{}_\nu = \hat{U} \widetilde{\Omega}^0_\nu \hat{U}^\dagger$. Given $\widetilde{\Omega}^0_\nu = U^{}_0 \cdot {\rm diag}\{\widetilde{m}^2_1, \widetilde{m}^2_2, \widetilde{m}^2_3\} \cdot U^{\dagger}_0$, one can get the final mixing matrix $\widetilde{U} = \hat{U} U^{}_0$. From the first part on the right-hand side of Eq.~(\ref{eq:omega0}), we can identify an intrinsic symmetry under
\begin{eqnarray}
    \theta^{}_{12} \to\theta^{}_{12} - \frac{\pi}{2}\; , ~~ m^{}_1 \leftrightarrow m^{}_2 \; ,
    \label{eq:symmetry}
    \end{eqnarray}
indicating $\{s^{}_{12}, c^{}_{12}\} \to \{-c^{}_{12}, s^{}_{12}\}$ and $\{\sin 2\theta^{}_{12}, \cos 2\theta^{}_{12}\} \to \{-\sin 2\theta^{}_{12}, -\cos 2\theta^{}_{12}\}$ for the mixing angle, and $\Delta^{}_{21} \to -\Delta^{}_{21}$ for the mass-squared difference. It is easy to verify that the effective Hamiltonian $\widetilde{H}^{}_{\rm eff}$ is invariant under these transformations. Note that if a different parametrization of $U$ is assumed, the transformations will be changed to those associated with the rightmost rotation matrix in $U$ and the corresponding mass eigenvalues.

One may argue that such a symmetry is spurious in the sense of just changing the parameter space from one part to another~\cite{deGouvea:2000pqg}. But this is not the case. To clearly see this point, we follow Ref.~\cite{deGouvea:2008nm} and discuss the physical ranges of $\theta^{}_{12}$ and $\Delta^{}_{21}$. First of all, there are two different ways to define neutrino mass eigenstates: (A) $\nu^{}_1$ is lighter than $\nu^{}_2$, i.e., $\Delta^{}_{21} > 0$; (B) $\nu^{}_1$ contains more component of $\nu^{}_e$, i.e., $|U^{}_{e1}|^2 = c^2_{12} > |U^{}_{e2}|^2 = s^2_{12}$. Then, one can determine the physical ranges of $\theta^{}_{12}$ and $\Delta^{}_{21}$: $\theta^{}_{12} \in [0, \pi/2]$ and $\Delta^{}_{21} > 0$ in Case (A); and $\theta^{}_{12} \in [0, \pi/4]$ and either $\Delta^{}_{21} > 0$ or $\Delta^{}_{21} < 0$ in Case (B), where $\theta^{}_{12} \to -\theta^{}_{12}$ can be compensated by redefining the phases of charged-lepton and neutrino fields. Moreover, as proved in Ref.~\cite{deGouvea:2008nm}, the points $(\theta^{}_{12}, \Delta^{}_{21})$ and $(\pi/2 - \theta^{}_{12}, \Delta^{}_{21})$ in Case (A) are equivalent to $(\theta^{}_{12}, \Delta^{}_{21})$ and $(\theta^{}_{12}, -\Delta^{}_{21})$ in Case (B). Therefore, the transformations in Eq.~(\ref{eq:symmetry}) and the equivalence between the parameter space in Case (A) and Case (B) can be summarized visually in a simple diagram
\begin{equation}
\centering
\xymatrix@C=4pc@R=4pc{
\left.(\theta^{}_{12}, \Delta^{}_{21})\right|^{}_{(A)}  \ar@{|=>}[r]^-{\theta^{}_{12} - \pi/2} & \left.(\pi/2 - \theta^{}_{12}, \Delta^{}_{21})\right|^{}_{(A)} \ar@{<=>}[d] \\
\left.(\theta^{}_{12}, \Delta^{}_{21})\right|^{}_{(B)} \ar@{<=>}[u] \ar@{<=|}[r]^-{m^{}_1 \leftrightarrow m^{}_2} & \left.(\theta^{}_{12}, -\Delta^{}_{21})\right|^{}_{(B)}}
\nonumber
\end{equation}
implying that the whole system should be invariant no matter which definition of neutrino mass eigenstates is taken.

For later convenience, we introduce a gauge parameter $\eta \in [0, 1]$ and separate an identity matrix from $\widetilde{\Omega}^0_\nu$, namely,
\begin{eqnarray}
\widetilde{\Omega}^0_\nu = \left[\eta m^2_1 + (1 - \eta) m^2_2\right] {\bf 1} + \left(\begin{matrix}  Ac^2_{13} + (\eta -c^2_{12}) \Delta^{}_{21} & \Delta^{}_{21}s^{}_{12}c^{}_{12} & A s^{}_{13} c^{}_{13} \cr \Delta^{}_{21}s^{}_{12} c^{}_{12} & (\eta -s^2_{12}) \Delta^{}_{21} & 0 \cr A s^{}_{13} c^{}_{13} & 0 & As^2_{13} + \Delta^{}_* \end{matrix}\right) \; ,
\label{eq:center}
\end{eqnarray}
where $\Delta^{}_* = \eta \Delta^{}_{31} + (1-\eta) \Delta^{}_{32}$. The definition of $\Delta^{}_*$ has been discussed by Parke~\cite{Parke:2016joa} and his collaborators~\cite{Minakata:2015gra,Denton:2016wmg}. In particular, it has been demonstrated that $\Delta^{}_{\rm c} \equiv c^2_{12} \Delta^{}_{31} + s^2_{12} \Delta^{}_{32}$ is more advantageous than any other combinations of $\Delta^{}_{31}$ and $\Delta^{}_{32}$ in description of reactor neutrino experiments~\cite{Parke:2016joa}. More recently, it has been found in Ref.~\cite{Li:2016pzm} that $\Delta^{}_{\rm c}$ can be implemented to greatly simplify the neutrino oscillation probabilities in matter, when the latter are expanded in terms of the small ratio $\alpha^{}_{\rm c} \equiv \Delta^{}_{21}/\Delta^{}_{\rm c} \approx 0.03$. However, the underlying reason for this simplification is not well justified in Ref.~\cite{Li:2016pzm}.

Now we have a closer look at the new form of $\widetilde{\Omega}^0_\nu$ in Eq.~(\ref{eq:center}). Since the effective Hamiltonian possesses an intrinsic symmetry under the transformations $\theta^{}_{12} \to \theta^{}_{12} - \pi/2$ and $m^{}_1 \leftrightarrow m^{}_2$ (i.e., $\Delta^{}_{21} \to -\Delta^{}_{21}$), it should also be respected by the manual separation in Eq.~(\ref{eq:center}). Retaining this symmetry in each part, we find only three solutions for $\eta$:
\begin{itemize}
\item {\it mean scheme} -- $\eta = 1/2$ and $\Delta^{}_{\rm m} \equiv \Delta^{}_*(\eta = 1/2) = (\Delta^{}_{31} + \Delta^{}_{32})/2$. In this scheme, we can obtain
\begin{eqnarray}
\widetilde{\Omega}^0_\nu = \frac{m^2_1 + m^2_2}{2} {\bf 1} + \Delta^{}_{\rm m} \left(\begin{matrix}  \widehat{A}^{}_{\rm m} c^2_{13} - \alpha^{}_{\rm m} c^{}_{2\theta^{}_{12}}/2 & \alpha^{}_{\rm m}s^{}_{2\theta^{}_{12}}/2 & \widehat{A}^{}_{\rm m} s^{}_{13} c^{}_{13} \cr \alpha^{}_{\rm m}s^{}_{2\theta^{}_{12}}/2 & \alpha^{}_{\rm m}c^{}_{2\theta^{}_{12}}/2 & 0 \cr \widehat{A}^{}_{\rm m} s^{}_{13} c^{}_{13} & 0 & \widehat{A}^{}_{\rm m}s^2_{13} + 1 \end{matrix}\right) \; ,
\label{eq:mean}
\end{eqnarray}
where $\widehat{A}^{}_{\rm m} \equiv A/\Delta^{}_{\rm m}$ and $\alpha^{}_{\rm m} \equiv \Delta^{}_{21}/\Delta^{}_{\rm m}$. This definition of $\Delta^{}_{\rm m}$ has already been used by the Bari group for a global-fit analysis of neutrino oscillation data~\cite{Capozzi:2016rtj}.

\item {\it cosine scheme} -- $\eta = c^2_{12}$ and $\Delta^{}_{\rm c} \equiv \Delta^{}_*(\eta = c^2_{12}) = c^2_{12} \Delta^{}_{31} + s^2_{12} \Delta^{}_{32}$. This choice has been adopted in a number of works by Parke and others~\cite{Parke:2016joa, Minakata:2015gra, Denton:2016wmg, Li:2016pzm}. For this scheme, we can get
\begin{eqnarray}
\widetilde{\Omega}^0_\nu = \left(m^2_1 c^2_{12} + m^2_2 s^2_{12}\right) {\bf 1} + \Delta^{}_{\rm c} \left(\begin{matrix}  \widehat{A}^{}_{\rm c} c^2_{13}  & \alpha^{}_{\rm c}s^{}_{2\theta^{}_{12}}/2 & \widehat{A}^{}_{\rm c} s^{}_{13} c^{}_{13} \cr \alpha^{}_{\rm c}s^{}_{2\theta^{}_{12}}/2 & \alpha^{}_{\rm c}c^{}_{2\theta^{}_{12}} & 0 \cr \widehat{A}^{}_{\rm c} s^{}_{13} c^{}_{13} & 0 & \widehat{A}^{}_{\rm c}s^2_{13} + 1 \end{matrix}\right) \; ,
\label{eq:cos}
\end{eqnarray}
where $\widehat{A}^{}_{\rm c} \equiv A/\Delta^{}_{\rm c}$ and $\alpha^{}_{\rm c} \equiv \Delta^{}_{21}/\Delta^{}_{\rm c}$. In the following two sections, we try to explain why the series expansions of oscillation probabilities in this scheme give us the most accurate and compact results.

\item {\it sine scheme} -- $\eta = s^2_{12}$ and $\Delta^{}_{\rm s} \equiv \Delta^{}_*(\eta = s^2_{12}) = s^2_{12} \Delta^{}_{31} + c^2_{12} \Delta^{}_{32}$. In this scheme, we can obtain
    \begin{eqnarray}
\widetilde{\Omega}^0_\nu = \left(m^2_1 s^2_{12} + m^2_2 c^2_{12}\right){\bf 1} + \Delta^{}_{\rm s} \left(\begin{matrix}  \widehat{A}^{}_{\rm s} c^2_{13} - \alpha^{}_{\rm s} c^{}_{2\theta^{}_{12}} & \alpha^{}_{\rm s}s^{}_{2\theta^{}_{12}}/2 & \widehat{A}^{}_{\rm s} s^{}_{13} c^{}_{13} \cr \alpha^{}_{\rm s}s^{}_{2\theta^{}_{12}}/2 & 0 & 0 \cr \widehat{A}^{}_{\rm s} s^{}_{13} c^{}_{13} & 0 & \widehat{A}^{}_{\rm s}s^2_{13} + 1 \end{matrix}\right) \; ,
\label{eq:sin}
\end{eqnarray}
where $\widehat{A}^{}_{\rm s} \equiv A/\Delta^{}_{\rm s}$ and $\alpha^{}_{\rm s} \equiv \Delta^{}_{21}/\Delta^{}_{\rm s}$. This definition has also been used for series expansions of neutrino oscillation probabilities that are numerically studied in Ref.~\cite{Li:2016pzm}.
\end{itemize}
Though all the formulas in Eqs.~(\ref{eq:mean}), (\ref{eq:cos}) and (\ref{eq:sin}) are equivalent to the original one in Eq.~(\ref{eq:omega0}), one can observe that each matrix element in $\widetilde{\Omega}^0_\nu$ in the symmetric formulation respects the symmetry indicated in Eq.~(\ref{eq:symmetry}). As a consequence, the parameters $\alpha$'s are now always combined with either $s^{}_{2\theta^{}_{12}} \equiv \sin 2\theta^{}_{12}$ or $c^{}_{2\theta^{}_{12}} \equiv \cos 2\theta^{}_{12}$ to form an invariant.

For comparison, we also explicitly write down $\widetilde{\Omega}^0_\nu$ in the normal scheme with $\eta = 1$, i.e.,
\begin{eqnarray}
\widetilde{\Omega}^0_\nu = m^2_1 {\bf 1} + \Delta^{}_{31} \left(\begin{matrix}  \widehat{A} c^2_{13} + \alpha s^2_{12} & \alpha s^{}_{2\theta^{}_{12}}/2 & \widehat{A} s^{}_{13} c^{}_{13} \cr \alpha s^{}_{2\theta^{}_{12}}/2 & \alpha c^2_{12} & 0 \cr \widehat{A}  s^{}_{13} c^{}_{13} & 0 & \widehat{A} s^2_{13} + 1 \end{matrix}\right) \; ,
\label{eq:normal}
\end{eqnarray}
where $\alpha \equiv \Delta^{}_{21}/\Delta^{}_{31}$ and $\widehat{A} \equiv A/\Delta^{}_{31}$ have been defined. It is straightforward to observe the relation $\Delta^{}_* = \Delta^{}_{31}\left[1 - (1 - \eta) \alpha\right]$ and the ``renormalization" of two important parameters
\begin{eqnarray}
\alpha^{}_* = \frac{\alpha}{1 - (1-\eta)\alpha} \; ,\quad \widehat{A}^{}_* = \frac{\widehat{A}}{1 - (1-\eta)\alpha} \; ,
\label{eq:renorm}
\end{eqnarray}
where the subscripts ``$\ast$" should be replaced by their counterparts in the symmetric schemes.

The eigenvalues of $\widetilde{\Omega}^{}_\nu$ can be calculated even without any specific parametrization of the PMNS matrix, and in a way independent of flavor basis~\cite{Barger,Zaglauer,Xing:2000gg}. However, here we are interested in the symmetric form in the standard parametrization of $U$, namely,
\begin{eqnarray}
\widetilde{m}^2_1 &=& \left[\eta m^2_1 + (1-\eta) m^2_2\right]  + \frac{1}{3} x - \frac{1}{3} \sqrt{x^2 - 3y} \left[z + \sqrt{3(1-z^2)}\right] \; , \nonumber \\
\widetilde{m}^2_2 &=& \left[\eta m^2_1 + (1-\eta) m^2_2\right] + \frac{1}{3} x - \frac{1}{3} \sqrt{x^2 - 3y} \left[z - \sqrt{3(1-z^2)}\right] \; , \nonumber \\
\widetilde{m}^2_3 &=& \left[\eta m^2_1 + (1-\eta) m^2_2\right] + \frac{1}{3} x + \frac{2}{3} z\sqrt{x^2 - 3y} \; ,
\label{eq:eigen}
\end{eqnarray}
where $x$, $y$ and $z$ are given by
\begin{eqnarray}
x &=& \Delta^{}_* \left[1 + \widehat{A}^{}_* + \left(2\eta - 1\right) \alpha^{}_*\right] \; , \nonumber \\
y &=& \Delta^2_* \left\{\widehat{A}^{}_* c^2_{13} + \frac{\alpha^{}_*}{2} \left[2(2\eta - 1) + \widehat{A}^{}_* c^{}_{2\theta^{}_{12}} c^2_{13} + \widehat{A}^{}_* (2\eta - 1)(1+s^2_{13})\right] + \eta(\eta - 1)\alpha^2_*\right\} \; , \nonumber \\
z &=& \cos \left\{\frac{1}{3} \arccos \frac{2x^3 - 9xy + 27 \alpha^{}_* \Delta^3_* \left[(\eta - s^2_{12})\widehat{A}^{}_*c^2_{13} + \eta(\eta - 1) (1+\widehat{A}^{}_*s^2_{13})\alpha^{}_*\right]}{2(x^2 - 3y)^{3/2}} \right\} \; . ~~~~~~~~
\label{eq:xyz}
\end{eqnarray}
It is worth mentioning that $x$ and $y$ depend on the gauge parameter $\eta$, whereas $x^2 - 3y$ and $z$ actually do not if they are expressed in terms of the original parameters $\Delta^{}_{31}$, $\alpha$ and $A$. The dependence on $\eta$ comes into play when we use $\Delta^{}_*$, $\alpha^{}_*$ and $\widehat{A}^{}_*$ and perform series expansions of the eigenvalues in terms of $\alpha^{}_*$.

\vspace{0.5cm}

{\bf Series expansions} --- It has been a longstanding problem in neutrino physics to derive more accurate and compact formulas for neutrino oscillation probabilities in matter, which could help explain the experimental results. One practically useful
approach is to expand the oscillation probabilities in terms of some small parameters, e.g., the ratio of two hierarchial neutrino mass-squared differences $\alpha \equiv \Delta^{}_{21}/\Delta^{}_{31} \approx 0.03$ and the smallest mixing angle $s^2_{13} \equiv \sin^2 \theta^{}_{13} \approx 0.02$ in the standard parametrization of $U$. See, e.g., Refs.~\cite{Cervera:2000kp,Freund:2001pn,Akhmedov:2004ny} for early development along this direction, and Ref.~\cite{Agarwalla:2013tza,Xu:2015kma,Minakata:2015gra,Flores:2015mah, Xing:2016ymg,Denton:2016wmg,Li:2016pzm} for recent progress.

For our purpose, it is instructive to concentrate first on two important functions $\sqrt{x^2 - 3y}$ and $z$ appearing in the mass eigenvalues in Eq.~(\ref{eq:eigen}). The exact formulas of them can be directly computed by using Eq.~(\ref{eq:xyz}), while their series expansions up to the second order of $\alpha^{}_*$ have been given in Ref.~\cite{Li:2016pzm}. To the first order of $\alpha^{}_*$, one can get
\begin{eqnarray}
z \approx \frac{1+\widehat{A}^{}_* + 3\widehat{C}^{}_*}{4\widehat{C}^\prime_*} &+& \frac{\alpha^{}_*}{4\widehat{C}^{}_* \widehat{C}^\prime_*} \left[2\widehat{C}^{}_* (1-2\eta) - 3(\eta - c^2_{12})(1 - \widehat{A}^{}_* c^{}_{2\theta^{}_{13}} - \widehat{C}^{}_*)\right] \nonumber \\
&-&\frac{\alpha^{}_* (1+\widehat{A}^{}_* + 3\widehat{C}^{}_*)}{8\widehat{C}^{\prime 3}_*} \left[ (1 - 2\eta)(1+\widehat{A}^{}_*) + 3\widehat{A}^{}_* c^2_{13} (\eta - c^2_{12})\right] \; ,
\label{eq:zexp}
\end{eqnarray}
and
\begin{eqnarray}
\sqrt{x^2 - 3y} \approx \Delta^{}_* \left\{\widehat{C}^\prime + \frac{\alpha^{}_*}{2\widehat{C}^\prime} \left[(1 - 2\eta) (1 + \widehat{A}^{}_*) + 3\widehat{A}^{}_* c^2_{13} (\eta - c^2_{12})\right] \right\} \; ,
\label{eq:xyexp}
\end{eqnarray}
where $\widehat{C}^{}_* \equiv [(1 - \widehat{A}^{}_*)^2 + 4\widehat{A}^{}_* s^2_{13}]^{1/2}$ and $\widehat{C}^\prime_* \equiv (\widehat{C}^2_* + \widehat{A}^{}_* c^2_{13})^{1/2}$ have been introduced. Some interesting observations are summarized below:
\begin{itemize}
\item Setting $\eta = 1/2$  or $\eta = c^2_{12}$, one can see that all the terms proportional to $1-2\eta$ or $\eta - c^2_{12}$ will disappear, leading to a great simplification of the approximate results in Eqs.~(\ref{eq:zexp}) and (\ref{eq:xyexp}). If we take another value $\eta = s^2_{12}$, both $1 - 2\eta$ and $\eta - c^2_{12}$ give the same factor $\cos 2\theta^{}_{12}$ up to a sign, so those two terms in the square brackets on the right-hand side of Eqs.~(\ref{eq:zexp}) and (\ref{eq:xyexp}) can be combined into a single one. In this sense, the choice of $\eta$ in all three symmetric schemes help derive simpler analytical results.

\item One can compute three eigenvalues to the first order of $\alpha^{}_*$ with the help of Eqs.~(\ref{eq:zexp}) and (\ref{eq:xyexp}). For illustration, we only quote the approximate result for $\widetilde{m}^2_3$ from Ref.~\cite{Li:2016pzm}
    \begin{eqnarray}
    \widetilde{m}^2_3 &\approx& m^2_2 -\eta\Delta^{}_{21} + \Delta^{}_* \left[\frac{1 + \widehat{A}^{}_* + \widehat{C}^{}_*}{2} - \frac{(\eta - c^2_{12})(1 - \widehat{C}^{}_* - \widehat{A}^{}_* c^{}_{2\theta^{}_{13}})}{2\widehat{C}^{}_*} \alpha^{}_* \right] \;,
    \label{eq:m3g}
    \end{eqnarray}
    which can reproduce the same result in Ref.~\cite{Freund:2001pn} by setting $\eta = 1$, namely,
    \begin{eqnarray}
    \widetilde{m}^2_3 \approx m^2_1 +  \Delta^{}_{31} \left[ \frac{1 + \widehat{A} + \widehat{C}}{2} + \frac{s^2_{12}(\widehat{C} - 1 + \widehat{A} c^{}_{2\theta^{}_{13}})}{2\widehat{C}} \alpha \right] \; .
    \label{eq:m3n}
    \end{eqnarray}
    On the other hand, in the cosine scheme with $\eta = c^2_{12}$, one can see the first-order term vanishes, and the leading-order contribution reads
    \begin{eqnarray}
    \widetilde{m}^2_3 \approx m^2_1 + s^2_{12} \Delta^{}_{21} + \Delta^{}_{\rm c} \frac{1 + \widehat{C}^{}_{\rm c} + \widehat{A}^{}_{\rm c}}{2} \; ,
    \label{eq:m3c}
    \end{eqnarray}
    where $\widehat{C}^{}_{\rm c} = [(1 - \widehat{A}^{}_{\rm c})^2 + 4\widehat{A}^{}_{\rm c} s^2_{13}]^{1/2}$ is implied. Therefore, higher-order terms start from ${\cal O}(\alpha^2_*)$ in the cosine scheme. In order to clarify that the leading-order result in Eq.~(\ref{eq:m3c}) is even more precise than that in Eq.~(\ref{eq:m3n}), we recall the definitions $\Delta^{}_{\rm c} \equiv \Delta^{}_{31} (1 - s^2_{12} \alpha)$ and $\widehat{A}^{}_{\rm c} \equiv \widehat{A}/(1 - s^2_{12} \alpha)$ and insert them into Eq.~(\ref{eq:m3c}). Expanding the function $\widehat{C}^{}_{\rm c}$ to the second order of $\alpha$, we arrive at
    \begin{eqnarray}
    \widetilde{m}^2_3 \approx m^2_1 &+& \Delta^{}_{31} \left[ \frac{1 + \widehat{A} + \widehat{C}}{2} + \frac{s^2_{12}(\widehat{C} - 1 + \widehat{A} c^{}_{2\theta^{}_{13}})}{2\widehat{C}} \alpha \right] \nonumber \\
    &+& \Delta^{}_{31}  \left[ \frac{s^4_{12}(\widehat{C} - 1 + \widehat{A} c^{}_{2\theta^{}_{13}}) (\widehat{C} + 1 - \widehat{A} c^{}_{2\theta^{}_{13}})}{4\widehat{C}^3}\alpha^2 + {\cal O}(\alpha^3) \right] \; ,
    \label{eq:m3ch}
    \end{eqnarray}
    which exactly reproduces the first-order result in Eq.~(\ref{eq:m3n}) and partly incorporates the second-order corrections. This can explain why the numerical precision in the cosine scheme is superior to that in the normal scheme, when the oscillation probabilities are expanded to the same order.
\end{itemize}
In a similar way, one can derive the results for $\eta = 1/2$ and $\eta = s^2_{12}$ and compare them with those in Eq.~(\ref{eq:m3ch}). Although the first-order terms are not vanishing in the mean and sine schemes, the final results involving the ``renormalized" parameters $\alpha^{}_{\rm m}$ and $\widehat{A}^{}_{\rm m}$ (or $\alpha^{}_{\rm s}$ and $\widehat{A}^{}_{\rm s}$) can also be regarded as a resummation of higher-order terms of $\alpha$. Since all three eigenvalues and oscillation probabilities have been given in Ref.~\cite{Li:2016pzm} for the general $\eta$ gauge, it is unnecessary to repeat them here.

\vspace{0.5cm}

{\bf Renormalization-group equations} --- Though we have seen that $\eta = c^2_{12}$ gives rise to the simplest results, as the first-order correction is vanishing, it is not understood why it should be so. From the symmetry arguments in the previous section, three schemes should be equally powerful in simplifying approximate formulas. In the following, we explain the reason by implementing the renormalztion-group equations (RGEs), which have been widely applied in quantum field theories~\cite{Stueckelberg:1953dz,GellMann:1954fq} and condense matter physics~\cite{Wilson:1971bg}. In our case, the central idea is that the exact mass eigenvalues of $\widetilde{H}^{}_{\rm eff}$ should be independent of the gauge parameter $\eta$. In fact, however, they are computed via perturbative expansions, and the dependence on $\eta$ actually comes in at any given order of $\alpha^{}_*$.

Assuming now $\eta$ to be an arbitrary positive parameter, which acts like the renormalization scale $\mu$ in relativistic quantum field theories, we shall examine the $\eta$-dependence of mass eigenvalues $\widetilde{m}^2_i$. First, as indicated in Eq.~(\ref{eq:renorm}), the exact dependence of $\alpha^{}_*$ and $\widehat{A}^{}_*$ on $\eta$ is already known, and can be reflected by the following RGEs
\begin{eqnarray}
\frac{{\rm d}\alpha^{}_*}{{\rm d}\eta} = -\alpha^2_* \; , \quad \quad \frac{{\rm d}\widehat{A}^{}_*}{{\rm d}\eta} = - \widehat{A}^{}_* \alpha^{}_* \; ,
\label{eq:RGEa}
\end{eqnarray}
where we have used the ``renormalized" parameters $\alpha^{}_*$ and $\widehat{A}^{}_*$ in the beta functions on the right-hand side of Eq.~(\ref{eq:RGEa}). Notice that these RGEs are the exact results, so we are actually dealing with an exactly solvable model. Then, it is easy to derive the RGE of $\widehat{C}^{}_*$ from its definition $\widehat{C}^2_* = (1 - \widehat{A}^{}_*)^2 + 4\widehat{A}^{}_* s^2_{13}$, i.e.,
\begin{eqnarray}
\frac{{\rm d}\widehat{C}^{}_*}{{\rm d}\eta} = \frac{\widehat{A}^{}_* - c^{}_{2\theta^{}_{13}}}{\widehat{C}^{}_*} \frac{{\rm d}\widehat{A}^{}_*}{{\rm d}\eta} = -\frac{\widehat{A}^{}_*(\widehat{A}^{}_* - c^{}_{2\theta^{}_{13}})}{\widehat{C}^{}_*} \alpha^{}_* \; .
\label{eq:RGEc}
\end{eqnarray}
The exact solutions to these RGEs are actually the definitions of $\alpha^{}_*$, $\widehat{A}^{}_*$ and $\widehat{C}^{}_*$ with $\alpha^{}_* = \alpha$, $\widehat{A}^{}_* = \widehat{A}$ and $\widehat{C}^{}_* = \widehat{C}$ at $\eta = 1$.

Second, the RGEs can be used to investigate the $\eta$-dependence of the eigenvalues $\widetilde{m}^2_i$. We take $\widetilde{m}^2_3$ for an illustrative example, and its approximate formula has been given in Eq.~(\ref{eq:m3g}). At the leading order of $\alpha^{}_*$, we calculate the derivative of $f^{(0)}(\eta) \equiv (\widetilde{m}^2_3 - m^2_2)/\Delta^{}_{31}$, where the superscript ``(0)" means that the zeroth-order term in $\widetilde{m}^2_3$ is included. The final result is
\begin{eqnarray}
\frac{{\rm d}}{{\rm d}\eta} f^{(0)}(\eta)  = \frac{1}{2} \left[1 - (1-\eta)\alpha\right] \left[ (\widehat{A}^{}_* + \widehat{C}^{}_* - 1)\alpha^{}_* + \frac{{\rm d}\widehat{A}^{}_*}{{\rm d}\eta} + \frac{{\rm d}\widehat{C}^{}_*}{{\rm d}\eta} \right] \; .
\end{eqnarray}
Requiring ${\rm d}f^{(0)}/{\rm d}\eta = 0$ and making use of the first identity in Eq.~(\ref{eq:RGEc}), one arrives at
\begin{eqnarray}
\frac{{\rm d}\widehat{A}^{}_*}{{\rm d}\eta} = - \frac{(\widehat{A}^{}_* + \widehat{C}^{}_* - 1)\widehat{C}^{}_*}{\widehat{A}^{}_* + \widehat{C}^{}_* - c^{}_{2\theta^{}_{13}}} \alpha^{}_* \; ,
\end{eqnarray}
which is different from the exact result of ${\rm d}\widehat{A}^{}_*/{\rm d}\eta$ in Eq.~(\ref{eq:RGEa}). This is reasonable because only the leading-order contribution is taken into account. Moreover, the RGE of $\alpha^{}_*$ is not involved at the leading order, which is also evident from its exact formula in Eq.~(\ref{eq:RGEa}).

Then we go to the first order of $\alpha^{}_*$, and define the function $f^{(1)}(\eta)\equiv (\widetilde{m}^2_3 - m^2_2)/\Delta^{}_{31}$, which now includes both leading- and first-order terms. After a quick calculation, we find
\begin{eqnarray}
\frac{{\rm d}}{{\rm d}\eta} f^{(1)}(\eta) &=& \frac{1}{2} \left[1 - (1-\eta)\alpha\right] \times \left\{ \left[(\widehat{A}^{}_* + \widehat{C}^{}_* - 1) - \frac{1 - \widehat{C}^{}_* - \widehat{A}^{}_* c^{}_{2\theta^{}_{13}}}{\widehat{C}^{}_*}\right] \alpha^{}_* \right. \nonumber \\
&& - \frac{(\eta -c^2_{12})(1 - \widehat{C}^{}_* - \widehat{A}^{}_* c^{}_{2\theta^{}_{13}})}{\widehat{C}^{}_*} \left(\alpha^2_* + \frac{{\rm d}\alpha^{}_*}{{\rm d}\eta}\right)
\label{eq:second} \\
&& \left. +  \frac{{\rm d}\widehat{A}^{}_*}{{\rm d}\eta} \left(1 + \frac{\eta - c^2_{12}}{\widehat{C}^{}_*}c^{}_{2\theta^{}_{13}} \alpha^{}_*\right) + \frac{{\rm d}\widehat{C}^{}_*}{{\rm d}\eta} \left[1 + \frac{(\eta - c^2_{12})(1 - \widehat{A}^{}_* c^{}_{2\theta^{}_{13}})}{\widehat{C}^2_*} \alpha^{}_*\right] \right\}\; . \nonumber
\end{eqnarray}
Inserting the exact RGEs of $\alpha^{}_*$, $\widehat{A}^{}_*$ and $\widehat{C}^{}_*$ from Eqs.~(\ref{eq:RGEa}) and (\ref{eq:RGEc}) into Eq.~(\ref{eq:second}), we obtain a considerably simple result
\begin{eqnarray}
\frac{{\rm d}}{{\rm d}\eta} f^{(1)}(\eta) = \frac{1}{2} \left[1 - (1-\eta)\alpha\right] (\eta - c^2_{12}) \widehat{A}^2_* s^2_{2\theta^{}_{13}} \alpha^2_*\; .
\label{eq:f1f}
\end{eqnarray}
This implies that the requirement for ${\rm d}f^{(1)}/{\rm d}\eta  = 0$ at the first order of $\alpha^{}_*$ is consistent with the exact RGEs of $\alpha^{}_*$, $\widehat{A}^{}_*$ and $\widehat{C}^{}_*$ for $\eta = c^2_{12}$, resembling the main feature of the exact formula of $\widetilde{m}^2_3$, i.e., ${\rm d}\widetilde{m}^2_3/{\rm d}\eta = 0$, in this case. Therefore, any higher-order contributions to the beta functions of $\widehat{A}^{}_*$ and $\alpha^{}_*$ will either vanish or be proportional to $(\eta - c^2_{12})^n$ with $n$ being a positive integer. For other different values of $\eta$, one should derive the RGEs of $\widehat{A}^{}_*$ and $\alpha^{}_*$ order by order until the exact results in Eq.~(\ref{eq:RGEa}) are reached. This observation gives a severe constraint on the structure of higher-order terms, and demonstrates that the perturbation results for the choice of $\eta = c^2_{12}$ are much simpler. It is interesting to apply this approach to other mass eigenvalues and also the oscillation probabilities.

A brief comparison between our findings with the existing results in Refs.~\cite{Parke:2016joa,Minakata:2015gra,Denton:2016wmg,Li:2016pzm} should be helpful. Although the advantages of $\eta = c^2_{12}$ in deriving compact and accurate formulas of neutrino oscillation probabilities have been emphasized in those works, it has not been observed that the underlying reason may be due to an intrinsic symmetry in the effective Hamiltonian and the $\eta$-dependence of higher-order terms in series expansions can be studied in a convenient way by implementing the RGE approach.

\vspace{0.5cm}

{\bf Summary} --- We have pointed out that the effective Hamiltonian for neutrino oscillations in matter possesses an intrinsic symmetry under the transformations $\theta^{}_{12} \to \theta^{}_{12} - \pi/2$ and $m^{}_1 \leftrightarrow m^{}_2$, if the standard parametrization of the PMNS matrix is adopted. Based on this symmetry, we suggest an introduction of the $\eta$-gauge neutrino mass-squared difference $\Delta^{}_* \equiv \eta \Delta^{}_{31} + (1-\eta) \Delta^{}_{32}$ and advocate three schemes with $\eta = 1/2$, $\eta = c^2_{12}$ and $\eta = s^2_{12}$, for which such a symmetry is respected at any order of perturbative expansions of $\alpha^{}_* \equiv \Delta^{}_{21}/\Delta^{}_*$. The expansion in terms of $\alpha^{}_*$ in such a symmetric formulation actually incorporates many higher-order terms of $\alpha$. This follows the spirit of resummation.

The effective Hamiltonian $\widetilde{H}^{}_{\rm eff}$ can be exactly solved for a constant matter density. In this exact formulation, the eigenvalues and the corresponding eigenvectors are independent of the gauge parameter $\eta$, so are the oscillation probabilities. It becomes important only when we calculate the physical quantities by using the perturbation theory, i.e., series expansions in terms of $\alpha^{}_*$. Therefore, a symmetric formulation does make sense.

We have shown that all three symmetric schemes are helpful in simplifying the analytical results, and provide a simple proof for $\eta = c^2_{12}$ as the best choice, following the idea of renormalization-group equations. Noticing that $\alpha^{}_{\rm c} = \alpha/(1-s^2_{12}\alpha)$ itself in the cosine scheme with $\eta = c^2_{12}$ can be expanded in terms of $\alpha$, and likewise for $\widehat{A}^{}_{\rm c} = \widehat{A}/(1 - s^2_{12}\alpha)$, we do expect that the numerical accuracy in this scheme is higher, as $s^2_{12} \approx 0.3$ is always appearing together with $\alpha$ and it is the smallest compared to its counterparts $0.5$ and $c^2_{12} \approx 0.7$ in other schemes. All these observations are instructive for understanding the phenomena of neutrino oscillations in matter and useful in practical calculations.

\section*{Acknowledgements}

The author is indebted to Yu-feng Li, Jue Zhang and Jing-yu Zhu for fruitful collaboration and intensive discussions on neutrino oscillations in matter, and to Prof. Zhi-zhong Xing for helpful comments. This work was supported in part by the National Recruitment Program for Young Professionals and by the CAS Center for Excellence in Particle Physics (CCEPP).

\end{document}